\newif\ifdraft\draftfalse
\DocumentMetadata{}
\documentclass[runningheads]{llncs}

\usepackage[type={CC},modifier={by},version={4.0},imagewidth=5em]{doclicense}
\usepackage[utf8]{inputenc}
\usepackage[T1]{fontenc}
\usepackage[maxbibnames=99,maxcitenames=2,backend=biber,natbib=true,style=numeric-comp]{biblatex}
\DeclareSourcemap{
  \maps[datatype=bibtex]{
    \map{
      \step[fieldsource=doi,final]
      \step[fieldset=url,null]
    }
    \map{
      \step[fieldset=editor,null]
      \step[fieldset=pages,null]
      \step[fieldset=volume,null]
      \step[fieldset=publisher,null]
      \step[fieldset=series,null]
    }
  }
}
\usepackage[svgnames]{xcolor}
\usepackage{hyperref}
\hypersetup{
    colorlinks,
    linkcolor={red!50!black},
    citecolor={blue!50!black},
    urlcolor={blue!80!black}
}

\usepackage[frozencache,cachedir=.]{minted}
\usepackage{graphicx}
\usepackage{amsmath}
\usepackage{cleveref}
\usepackage{enumitem}
\usepackage[misc,geometry]{ifsym}
\usepackage{booktabs}
\usepackage{soul}
\usepackage{stmaryrd,xspace}
\usepackage{tikz}
\usetikzlibrary{arrows,shapes,positioning,fit,calc,decorations.pathreplacing,shapes.misc,shapes.geometric,decorations.pathmorphing,tikzmark,chains,tikzmark,scopes,matrix,backgrounds,patterns,shapes.multipart}
\definecolor{OwlRed}{RGB}{255,92,168}
\definecolor{OwlGreen}{RGB}{90,168,0}
\definecolor{OwlBlue}{RGB}{0,152,233}
\definecolor{OwlYellow}{RGB}{242,147,24}
\colorlet{OwlCyan}{OwlGreen!50!OwlBlue}
\colorlet{OwlOrange}{OwlRed!50!OwlYellow}
\colorlet{OwlBrown}{OwlRed!50!OwlGreen}
\colorlet{OwlViolet}{OwlRed!50!OwlBlue}
\colorlet{Py}{OwlCyan}%
\colorlet{C}{OwlOrange}%
\colorlet{Uni}{OwlBlue}%

\tikzset{
  invisible/.style={opacity=0},
  visible on/.style={alt=#1{}{invisible}},
  alt/.code args={<#1>#2#3}{%
    \alt<#1>{\pgfkeysalso{#2}}{\pgfkeysalso{#3}} %
  },
}

\tikzstyle{dom} = [draw, rectangle, rounded corners, inner sep = 3pt,font=\footnotesize]
\tikzstyle{combiner} = [draw, circle, inner sep = 0pt, minimum size=15pt, font=\footnotesize]
\tikzstyle{iterator} = [dom,fill=MidnightBlue!40]
\tikzstyle{memory} = [dom,fill=JungleGreen!40]
\tikzstyle{scalar} = [dom,fill=Tan!40]
\tikzstyle{numeric} = [dom,fill=Orchid!40]
\tikzstyle{lib} = [dom,fill=BrickRed!40]
\tikzstyle{shareddomain} = [dom,fill=OwlBlue!50]
\tikzstyle{optshareddomain} = [dom,dotted,fill=OwlBlue!50]
\tikzstyle{clangdomain} = [dom,fill=OwlOrange!50]
\tikzstyle{optclangdomain} = [dom,dotted,fill=OwlOrange!50]
\tikzstyle{pylangdomain} = [dom,fill=OwlCyan!50]
\tikzstyle{analysisdomain} = [dom,fill=OwlRed!50]

\ifdraft
  \usepackage{todonotes}
  \usepackage{luatex85}
  \pdfpagesattr{/CropBox [92 100 523 740]} %
\else
  \usepackage[disable]{todonotes}
\fi

\usepackage{multirow}

\usepackage{orcidlink}

\usepackage{blindtext}                      %
\usepackage{graphicx}                       %
\usepackage[firstpage]{draftwatermark}      %

\title{Try-Mopsa: \\Relational Static Analysis in Your Pocket\thanks{This work is partially supported by grant agreement ANR-24-CE25-7956-01 RAISIN from the French Agence Nationale de la Recherche, and by an Amazon Research Award, Fall 2024.}}

\author{Raphaël Monat\inst{1}\textsuperscript {(\href{mailto:raphael.monat@inria.fr}{\Letter})}\orcidlink{0000-0001-8487-0326}}
\authorrunning{R. Monat}

\institute{
  Univ. Lille, Inria, CNRS, Centrale Lille, \\UMR 9189 CRIStAL, F-59000 Lille, France
}

\begin{document}
\maketitle

\SetWatermarkAngle{0}
\SetWatermarkText{\raisebox{14.5cm}{%
  \hspace{10.3cm}%
\href{http://doi.org/10.5281/zenodo.17157478}{\includegraphics[width=20mm,keepaspectratio]{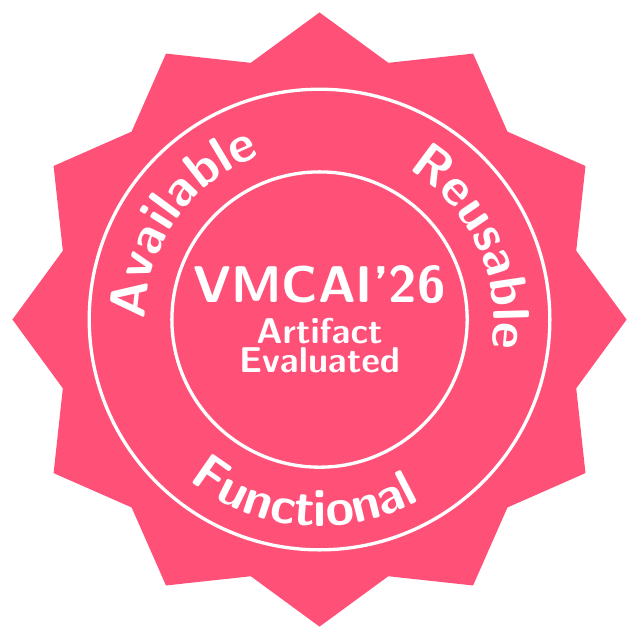}}%
}}

\begin{abstract}
  Static analyzers are complex pieces of software with large dependencies.
  They can be difficult to install, which hinders adoption and creates barriers for students learning static analysis.
  This work introduces Try-Mopsa: a scaled-down version of the Mopsa static analysis platform, compiled into JavaScript to run purely as a client-side application in web browsers.
  Try-Mopsa provides a responsive interface that works on both desktop and mobile devices.
  Try-Mopsa features all the core components of Mopsa.
  In particular, it supports relational numerical domains.
  We present the interface, changes and adaptations required to have a pure JavaScript version of Mopsa.
  We envision Try-Mopsa as a convenient platform for onboarding or teaching purposes. %

  \keywords{Static Analysis \and Abstract Interpretation \and Usability \and Teaching.}
\end{abstract}

\section{Introduction}

Static analyzers are complex pieces of software, usually building on a large number of dependencies.
For example, the Mopsa static analysis platform \cite{DBLP:conf/vstte/JournaultMMO19} requires among others: two parsing libraries (Menhir and libclang), the Zarith library to handle arbitrary precision arithmetic, and the Apron library to handle relational numerical domains.
When facing a large number of users (or students), there is always a chance to encounter some installation issues.
While good packaging or containerization can certainly limit those, installing a new tool still consumes time and resources.
In any case, the installation process hinders both testing and adoption of new static analysis tools.

One remedy to this issue is to provide zero-install ways to try a software, for example by enabling tool usage through a web browser.
This article presents Try-Mopsa, a scaled-down version of the Mopsa static analysis platform that runs entirely as a client-side web application.
It relies on a responsive interface to support a wide variety of devices (from smartphones to computers), and supports all core features of Mopsa, including relational domains and an interactive engine acting as an abstract debugger.
Try-Mopsa is available online \cite{try-mopsa-online}.
Thanks to its purely client-side implementation, Try-Mopsa is unaffected by the scalability issues of server-side implementations serving a large number of concurrent users.

We provide a brief overview of Mopsa in \Cref{sec:mopsa}.
Then, we describe in \Cref{sec:adaptation} how we adapted Mopsa to make it runnable in a web page.
\Cref{sec:interface} provides an overview of the resulting web interface,
\Cref{sec:di} evaluates several features of the implementation, and \Cref{sec:rw} discusses related work.

\section{A Brief Overview of Mopsa}
\label{sec:mopsa}

Mopsa is a Modular Open Platform for Static Analysis, rooted within the abstract interpretation framework \cite{CC77}.
It aims at providing a convenient platform for static analysis learners, developers and users.
Although Mopsa explores some new perspectives for the design of static analyzers, it is stable and precise enough to be on-par with state-of-the-art academic program analyzers participating to the Software-Verification Competition \cite{DBLP:conf/tacas/Beyer23,DBLP:conf/tacas/MonatOM23}.
\citet{DBLP:conf/vstte/JournaultMMO19} describe the core of Mopsa's principles, and \citet[Chapter 3]{mathese} provides an in-depth introduction to Mopsa's architecture.
We briefly describe three features of interest for this article:
\begin{description}
\item[Multilanguage support.] Mopsa supports the analysis of multiple programming languages. Currently, it supports the analysis of an in-house toy imperative language (called ``Universal''), of C \cite{DBLP:conf/sas/OuadjaoutM20} and of Python \cite{DBLP:conf/ecoop/MonatOM19,DBLP:conf/sas/MonatOM21}.

\item[User-defined analysis combination.] As an analysis platform, Mopsa offers a wide variety of abstract domains to choose from. Users define in a configuration file which abstract domains they want to enable, and how they should be combined (reduced product, ...).

\item[Tailored for relational domains.] Relational domains greatly improve the expressiveness of analyses by being able to infer constraints between variables.
  In addition, every abstract domain can introduce ghost variables and add constraints on those, which can be handled by an underlying relational domain.
\end{description}

\begin{figure}[!t]
  \centering
  \begin{tikzpicture}[node distance=.7cm and .7cm]
    \tikzstyle{block} = [rectangle, draw, text width=7em, text centered, minimum height=1.5em]
    \tikzstyle{label} = [midway,scale=0.7,rectangle,
    text width=6em, text centered]
    \tikzstyle{alternatives} = [rectangle split, rectangle split parts=#1,draw,rectangle split horizontal, rectangle split part align=base]

    \node (prog) { Program };
    \node[right=of prog] (config) { Configuration };
    \node[right=of config] (options) { Options };

    \node[below=of prog] (frontendt) { \phantom{y}Frontend\phantom{y} };
    \node[alternatives=3,below =.5em of frontendt,rectangle split part fill={white,white,gray!75}] (frontend-parts) {Universal\nodepart{two}Python\nodepart{three}C};
    \node [draw=black, fit={(frontendt) (frontend-parts)}] (frontend) {};

    \node[right=5em of frontendt] (analysisbuildert) { Analysis Builder };
    \node[below =.5em of analysisbuildert,draw=none] (analysisbuilder-parts) {\phantom{Universal Python C}};
    \node [draw=black, fit={(analysisbuildert) (analysisbuilder-parts)}] (analysisbuilder) {};

    \node[below = of $(frontend.south)!0.5!(analysisbuilder.south)$] (enginet) { Toplevel Analysis };
    \node[block,below= of enginet,xshift=-5em,yshift=1em] (it) {Iterators};
    \node[block,below= of enginet,xshift=+5em,yshift=1em] (part) {Partitioning};
    \node[block,below= of it] (itv) {Int Intervals};
    \node[block,right= of itv] (congr) {Congruences};
    \node[block,below= of itv,fill=gray!75] (fitv) {Float Intervals};
    \node[block,right= of fitv] (sp) {String Powerset};

    \node[below=of fitv,xshift=1em] (relt) {Relational Domain};
    \node[alternatives=2,below=.5em of relt,rectangle split part fill={gray!75,white}] (rel-parts) {Apron\nodepart{two}VPL};
    \node [draw=black, fit={(relt) (rel-parts)}] (rel) {};

    \node[block,xshift=2.5em,below=of sp,yshift=.4em,text width=2em] (dots) {$\ldots$};

    \node [draw=black, fit={(enginet) (it) (part) (itv) (congr) (fitv) (sp) (dots) (rel-parts)},inner sep=.75em] (engine) {};

    \node[below=4em of engine] (interfacet) { User Interface };
    \node[alternatives=3,below=.5em of interfacet,rectangle split part fill={white,white,gray!75}] (engine-parts) {Automatic\nodepart{two}Interactive\nodepart{three}DAP};
    \node [draw=black, fit={(interfacet) (engine-parts)},inner sep=.75em] (interface) {};

    \draw[->] (prog) -- (frontend);
    \draw[->] (config) -- (analysisbuilder);
    \draw[->] (options) -- (analysisbuilder);

    \draw[->] (frontend) -- (engine);
    \draw[->] (analysisbuilder) -- (engine);

    \draw[->] ([xshift=1.5em]engine.south) to [bend left] ([xshift=1.5em]interface.north);
    \draw[->] ([xshift=-2em]interface.north) to [bend left] ([xshift=-1.5em]engine.south);

  \end{tikzpicture}
  \caption{Components of Mopsa. All components are supported by Try-Mopsa, except those filled in gray.}
  \label{fig:mopsa}
\end{figure}
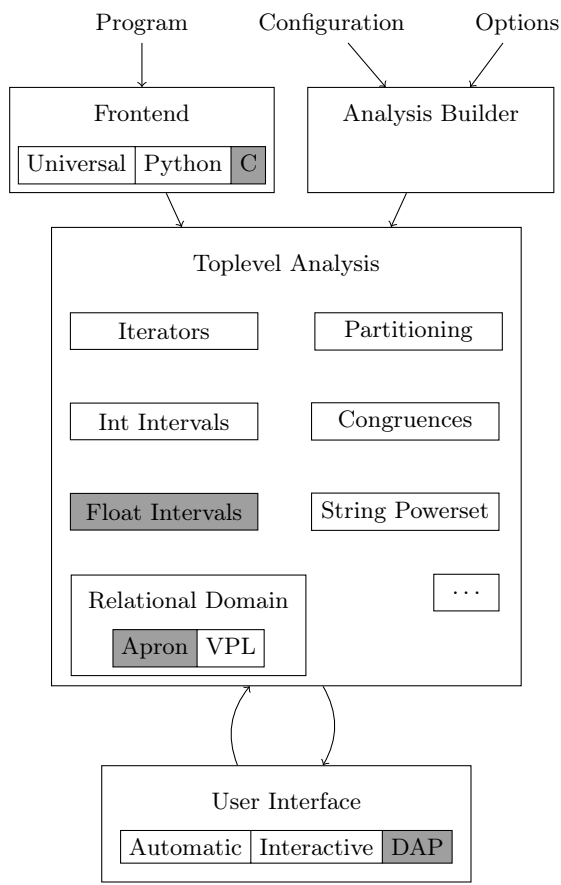

We show in \Cref{fig:mopsa} the main components of Mopsa, and describe them below.
All components are supported by Try-Mopsa, except those filled in gray.
We discuss unsupported components and potential replacements in \Cref{sec:adaptation}.

The frontend handles the parsing of a program into an abstract syntax tree (AST).
As we mentioned above, Mopsa currently supports the analysis of three programming languages.

The analysis builder takes a JSON configuration file describing the choice of abstract domains and their combinators, makes sure it is valid, and enables the corresponding abstract domains in the toplevel analysis.
It also sets passed options (either for the framework or for the enabled abstract domains).

Once the parsing is done and the abstract domains are combined according to the specified configuration, the toplevel analysis starts.
It runs the combination of chosen abstract domains, such as iterators handling loops and function calls, trace and state partitioning, numerical abstract domains and string abstractions.

Different user interfaces are offered by Mopsa.
The automatic interface is the classic one: the analysis runs to completion and then displays the results.
The interactive engine lets user navigate the abstract execution of the program, where analysis computations are performed on-the-fly accordingly.
This interface acts as a \texttt{gdb}-like abstract debugger, and supports breakpoints, program navigation, printing of abstract states, ...
The DAP interface provides a debug adapter protocol interface, similar to the interactive engine but allowing interactions with IDEs supporting this protocol.
\citet[Section 5]{DBLP:journals/sttt/MonatOM24} describe the former two interfaces in more details.

\section{Try-Mopsa: Under the Hood}%
\label{sec:adaptation}

This section describes the implementation of Try-Mopsa.
\Cref{sec:js} focuses on the backend part, explaining how we adapted Mopsa to be able to compile it to JavaScript.
\Cref{sec:wi} shows how this compiled JavaScript is integrated within a web page to provide a user-friendly interface.

\subsection{Compiling Mopsa to JavaScript}
\label{sec:js}

According to \texttt{cloc} \cite{adanial_cloc}, the current version of Mopsa consists of 87,856 lines of code. The overwhelming majority of the codebase (89\%) is written in OCaml.
We rely on the Js\_of\_ocaml (JSOO) compiler \cite{DBLP:journals/spe/VouillonB14} to compile these components of Mopsa to JavaScript.
This compiler option is selected in the build system as a drop-in replacement to the standard OCaml compiler creating executables.
This compilation process already supports most Mopsa features, including the parsing of user-defined analysis combination, fixpoint and interprocedural iterators, heap and string abstractions, trace and state partitioning.

However, some crucial features of Mopsa rely on external dependencies: we use Menhir or libclang to parse programs, the Zarith library to handle arbitrary precision arithmetic, and the Apron library to handle relational numerical domains.
We now discuss how these components have been adapted to compile to JavaScript.

\paragraph{Parsing libraries.}
Mopsa relies on the Menhir library \cite{menhir} to implement parsers for the Universal and the Python programming languages.
As Menhir is written in pure OCaml, JSOO natively supports compiling it to JavaScript.

Try-Mopsa currently does not support C parsing.
Indeed, Mopsa depends on LLVM and its libclang library to parse C programs, and our C parser contains 4,182 lines of C++ glue code.
While JSOO can interface with manually written JavaScript stubs, and some prototype versions of LLVM have been compiled to WebAssembly, we believe the current level of support for these processes would require too much manual work.

\paragraph{Arbitrary-precision integer arithmetic.}
Our integer abstractions (intervals, congruences) rely on arbitrary-precision arithmetic to avoid overflows and unsoundness.
Mopsa relies on the Zarith \cite{zarith} library to implement these mathematical integers.
Try-Mopsa makes use of a third-party alternative implementation in JavaScript of the Zarith interface, called Zarith\_stubs\_js \cite{zarith_stubs_js} and natively supported by the compilation process of JSOO.

\paragraph{Relational Domains.}
Mopsa depends on the Apron library \cite{DBLP:conf/cav/JeannetM09} to support relational numeric domains such as octagons \cite{octagons} and polyhedra \cite{polyhedra}.
However, Apron abstract domains are written in C/C++: compiling them to JavaScript, just like our C frontend, would be quite cumbersome.
For Try-Mopsa, we chose to replace Apron with the alternative Verified Polyhedra Library (VPL) \cite{DBLP:conf/synasc/BoulmeMMPY18}, which provides different implementations of polyhedra.
Given that some domains of the VPL are pure OCaml implementations, JSOO is able to compile it to JavaScript, enabling relational abstract domains to run within web browsers.
We wrote a new VPL wrapper, making it compatible with Mopsa's interface, and adding support for the \texttt{fold/expand} operators \cite{gopan} required by Mopsa but not built in the VPL.

\paragraph{Floating-point Abstractions.}
Our current implementation of floating-point intervals sets various rounding modes in the floating-point unit to be sound \cite{DBLP:conf/esop/Mine04}.
Since this is not supported by JavaScript, these intervals are currently not supported by Try-Mopsa.
Alternative implementations of floating-point intervals, not relying on specific rounding modes could be developed in future work to alleviate this limitation.

\subsection{Handling Web Interactions}
\label{sec:wi}

Try-Mopsa is split between a toplevel, handling the page interaction with the user and a web worker, handling the actual computations of the analysis.
Thanks to this decoupling, the web worker and the page rendering lie within different threads of the web browser, so longer computations do not freeze the browser's rendering process.

\paragraph{Web Worker.} The web worker relies primarily on the JavaScript executable obtained through the process described in the previous section.

The compiled version of Mopsa relies on standard output to interact with the user.
Additionally, the interactive engine reads user commands from the standard input.
We rely on JSOO's utilities to set up callbacks intercepting any standard input/output and redirecting them to the toplevel.
Thanks to this approach, Mopsa did not require any modification to work in the browser.

We rely on the virtual filesystem mechanism of JSOO to embed internal files Mopsa may expect to find during its analysis.
This currently concerns Python stubs for various library modules.

\paragraph{Toplevel.} The toplevel renders the initial dynamic elements of the page and updates them in response to user action.

It uses the Ace editor \cite{ace}, and the Ezjs\_ace bindings \cite{ezjs_ace} to display the input program and the analysis results.
We have extended the Ace editor to provide syntax highlighting for the Universal language, to allow collapsing of printed elements within the analysis result, and to render ANSI escape codes in HTML, as those are used by Mopsa to print in color.

The toplevel also draws a form so that user can select options.
The form is automatically generated from the option metadata encoded within Mopsa (which are fetched through a query to the web worker).
The options depend on which abstract domains are activated, and thus on the configuration chosen by the user.
The toplevel ensures that the option form is redrawn whenever the configuration is changed.

When a user starts an analysis, the toplevel creates a web worker to perform it.
The user may choose to interrupt the analysis, in which case the toplevel then signals the web worker to stop.

\paragraph{Interactive engine.}
The interactive engine of Mopsa follows an interaction loop by awaiting user input (in a synchronous, blocking fashion) and providing corresponding results.
However, web interfaces typically rely on asynchronous processes.
We explored options such as implementing an asynchronous interactive engine.
This would have required the decoupling of the interactive interface from the analysis computations, which would have been a large implementation effort introducing breaking changes.
In the end, we chose to rely on the sync-message \cite{sync-message} JavaScript library enabling synchronous communication between the web worker to the toplevel, should some user input be required.
Thanks to this embedding, all user-interface features of Mopsa are supported in Try-Mopsa.

\section{Try-Mopsa: Interface Overview}
\label{sec:interface}

\begin{figure}[!pt]
  \includegraphics[width=\textwidth]{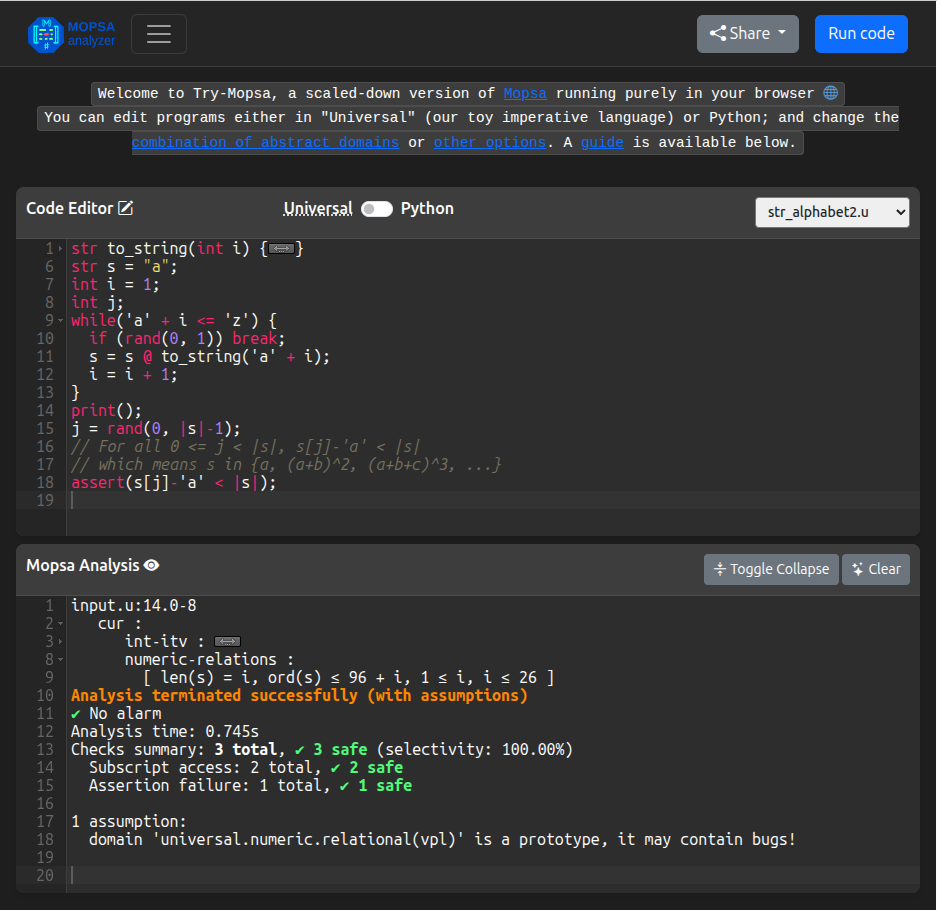}
  \vspace{-2em}
  \caption{Interface of Try-Mopsa: landing page, with program editor and analysis output.}
  \label{fig:itf:editor}
  \begin{minted}[escapeinside=€€,numbers=left,xleftmargin=2em,xrightmargin=0em,autogobble]{c}
€\PYG{k+kt}{str}€ to_string(int i) { /* omitted */ } €\setcounter{FancyVerbLine}{5}€
€\PYG{k+kt}{str}€ s = "a";
int i = 1;
int j;
while('a' + i <= 'z') {
  if (€\textcolor{blue}{\texttt{rand}}€(0, 1)) break;
  s = s €@€ €\textcolor{blue}{\texttt{to\_string}}€('a' + i);
  i = i + 1;
}
€\textcolor{blue}{\texttt{print}}€();
j = €\textcolor{blue}{\texttt{rand}}€(0, |s|-1);
// For all 0 <= j < |s|, s[j]-'a' < |s|
// which means s in {a, (a+b)^2, (a+b+c)^3, ...}
€\textcolor{blue}{\texttt{assert}}€(s[j]-'a' < |s|);
\end{minted}
  \vspace{-1.5em}
\caption{Program \texttt{str\_alphabet2.u} used as example.}
\label{fig:alphabet2}
  \vspace{-.5em}
\end{figure}

\begin{figure}[!t]
  \includegraphics[width=\textwidth]{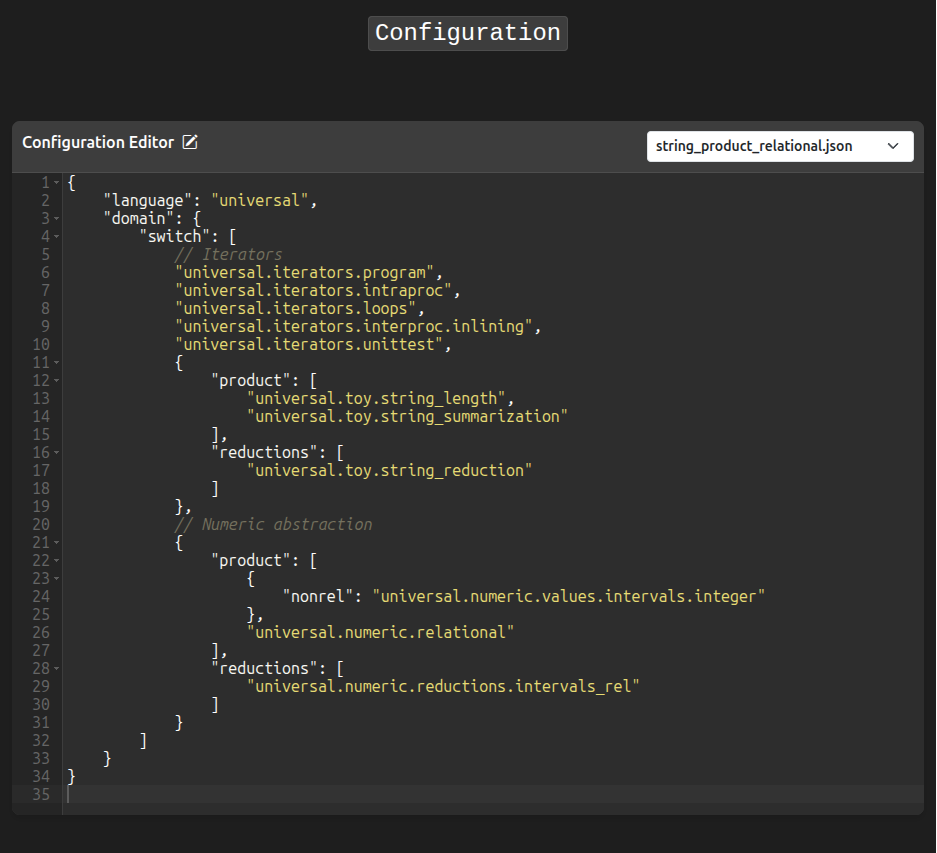}
  \caption{Interface of Try-Mopsa: configuration editor, allowing to specify the choice of abstract domains and their combinators.}
  \label{fig:itf:config}
\end{figure}

\begin{figure}[!t]
  \includegraphics[width=\textwidth]{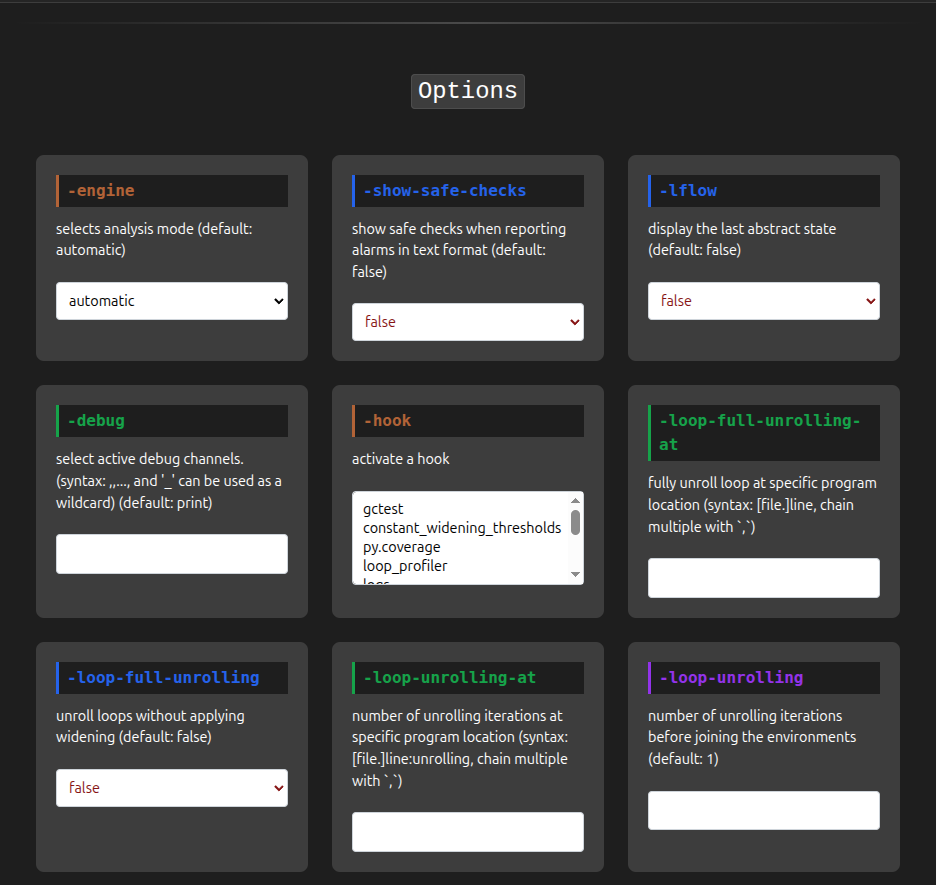}
  \caption{Interface of Try-Mopsa: option selection.}
  \label{fig:itf:options}
\end{figure}

The interface of Try-Mopsa is displayed in \Cref{fig:itf:editor,fig:itf:config,fig:itf:options} and can be tried online \cite{try-mopsa-online}.
The interface is responsive, to ensure it can be used on a wide variety of
screens, from smartphones to desktops.

The landing page shown in \Cref{fig:itf:editor} provides a code editor and displays analysis results.
By default, input programs are written in our toy imperative language.
Users can also switch to the Python analysis if they wish to, and they can load some example programs.
The analysis output consists of abstract states displayed when the \texttt{print()} instruction is analyzed, as well as a report on the runtime errors potentially detected by the analyzer.
If the interactive engine has been enabled, the interaction happens in this same box.
Note that on wider screens, the responsive design displays the code editor and analysis results side-by-side.

In \Cref{fig:itf:editor}, we loaded the example program \texttt{str\_alphabet2.u}, reproduced textually in \Cref{fig:alphabet2}.
Note that in the web editor, the definition of \texttt{to\_string}, acting as a cast, is folded to highlight the editor capabilities.

The program starts with the string \texttt{s}, initialized to the letter "a".
It then runs a loop until the number defined by \mintinline{c}{'a' + i} reaches the end of the lowercase alphabet.
Similarly to C, in Universal, characters (between single quotes) are interpreted as integers through their ASCII code.
This loop may stop non-deterministically at any iteration (line 10).
At each loop iteration, the string \texttt{s} is concatenated (with the @ operator) with the singleton string whose character corresponds to the integer \mintinline{c}{'a' + i}.
Thus the program non-deterministically computes a string \texttt{s} among $S = \{\text{"}a\text{"}, \text{"}ab\text{"}, \text{"}abc\text{"}, \ldots, \text{"}abc\cdots{}xyz\text{"}\}$.
The assertion at line 18 checks a simpler property: that \texttt{s} belongs to $S$ but up to any permutation of the characters.

\Cref{fig:itf:config} shows the configuration editor, letting users specify their choice of abstract domains and combinators.
Similarly to the code editor, users can also load a configuration from a pre-defined list.
We comment on the loaded configuration, \texttt{string\_product\_relational.json}.
This configuration features three notable abstract domains, working together.
The string length domain handles ghost variables representing the length of each string.
The string summarization handles ghost variables representing a summary of the contents of each string, through the ASCII codes of the contained characters.
These two domains are defined as a reduced product as they infer complementary pieces of information \cite{mathese}.
Finally, a relational numerical domain is enabled.
It handles the constraints passed by the aforementioned string domains, which allows it to infer relations between those different ghost variables, and program variables.

The result of analyzing \texttt{str\_alphabet2.u} using \texttt{string\_product\_relation-\allowbreak{}al.json} is displayed at the bottom of \Cref{fig:itf:editor}.
In this case, Try-Mopsa is able to prove the complex assertion at line 18, which mixes information about the contents and the length of string \texttt{s}.
Let us briefly comment on the abstract state inferred after the loop and printed at lines 1-9 of the analysis result in \Cref{fig:itf:editor}.
We can see that the relational domain inferred equality between the length of the string \texttt{s} and the integer variable \texttt{i}.
We can further notice an interesting relation between the contents of \texttt{s} ($\text{ord}(s)$) and its length: $\text{ord}(s) \leq \text{'}a\text{'} + \text{len}(s) - 1$.
This invariant has been obtained thanks to the cooperation between the string abstract domains and the relational abstract domain.
Note that the second inequality is exactly the invariant proved by Mopsa at line 18.

\Cref{fig:itf:options} displays various options to tweak the analysis.
As we mentioned in the previous section, these options are dynamically generated by the toplevel, depending on which abstract domains are activated.
In this example, the first five options are defined at the framework level, while the last four options customize the behavior of the loop iterator.
Note that the option's argument can be selected through different interfaces depending on the expected output (e.g., choice among fixed list, integer, string).

To further improve its usability, Try-Mopsa can also be used as a progressive web app, meaning that it can be installed locally on a device and run without relying on the network.
In addition, a share button (in the top-right corner of \Cref{fig:itf:editor}) lets users share their current program, configuration and option choices by encoding it into a URL.
This can be useful for example to simplify practical sessions where no manual loading is required.

\section{Implementation Discussion}
\label{sec:di}

This section briefly discusses the implementation size, maintainability, performance and browser compatibility of Try-Mopsa.

\paragraph{Implementation.}
The webpage, including extensions of the Ace editor for our purposes, is written in around 1,500 lines of HTML, CSS and JavaScript.
The toplevel and worker of Try-Mopsa consist of 670 lines of OCaml code.
The implementation within Mopsa of the VPL binder adds around 500 lines of OCaml code.
Once compiled with full optimizations, the worker and all the Mopsa code relying on it are compiled into a relatively compact JavaScript file of 3.1 megabytes.

\paragraph{Maintainability.}
We argue that the implementation of Try-Mopsa is maintainable with respect to the standard implementation of Mopsa.
Indeed, Try-Mopsa does not introduce breaking changes, meaning future development can be shared in the same repository.

In addition, most components of Mopsa were reused as-is, thanks in particular to the reuse of the same input-output loop as a terminal user of Mopsa would.
The only major component that has been added is the support for relational abstract domains through the VPL library.
This implementation provides adequate results on the toy examples we provide along with Try-Mopsa.

\begin{table}[!t]
  \centering
  \caption{Preliminary evaluation of analysis times with relational abstract domains enabled. The mean and standard deviation have been computed over 10 runs.}
  \begin{tabular}{lll}
    \toprule
    Program\phantom{aaaaaaaaa}          & Binary execution \phantom{aa} & Firefox execution\\
    \midrule
    attributes.py    & 0.09s $\pm$ 0.00 & 0.52s $\pm$ 0.03  \\
    fspath.py        & 0.09s $\pm$ 0.00 & 0.49s $\pm$ 0.02  \\
    list.py          & 0.15s $\pm$ 0.01 & 0.88s $\pm$ 0.03  \\
    loop.py          & 0.12s $\pm$ 0.01 & 0.72s $\pm$ 0.03  \\
    recency.py       & 0.08s $\pm$ 0.01 & 0.63s $\pm$ 0.03  \\
    \midrule
    str\_alphabet.u   & 0.04s $\pm$ 0.01 & 0.20s $\pm$ 0.01  \\
    str\_alphabet2.u  & 0.23s $\pm$ 0.01 & 0.81s $\pm$ 0.01  \\
    str\_conc\_loop.u  & 0.08s $\pm$ 0.01 & 0.29s $\pm$ 0.01  \\
    str\_conc\_loop2.u & 0.04s $\pm$ 0.01 & 0.15s $\pm$ 0.01  \\
    \bottomrule
  \end{tabular}
  \label{tbl:runtimes}
\end{table}

\paragraph{Performance.}
  The goal of Try-Mopsa is to ease tool demonstrations and testing without installation, where raw analysis performance is not a priority.
  Nevertheless, we ran a preliminary evaluation comparing the analysis times between the natively generated executable, and the same code running in Firefox.
  We ran this comparison on the toy, default examples of Try-Mopsa, with configurations using relational abstract domains.
  We show in \Cref{tbl:runtimes} the nine most significant programs to analyze.
  On average, Try-Mopsa is five times slower than native binary execution.
  While this is a noticeable slowdown, we believe the analysis times -- below one second -- make Try-Mopsa still usable for demonstration or teaching purposes.
  Future work could focus on using JSOO's recent WebAssembly code generation features to obtain more efficient code.

\paragraph{Browser compatibility.}
We use the Playwright framework \cite{playwright} to test the browser compatibility of Try-Mopsa.
The tests run multiple usage scenarios and check the results are visible and as expected.
These tests run during our continuous integration process and have helped identify several usability issues, particularly on mobile browsers.
We currently test three desktop browsers (Chrome, Firefox, Safari), and two mobile browsers (Chrome, Safari), using different viewports corresponding to popular iPhone and Android devices, either used in portrait or in landscape orientation.
There are currently 5 test scenarios, resulting in 35 tests depending on the chosen browser and viewport.
Additional tests ensure that all examples programs can be analyzed in the relevant example configurations proposed in our interface.
We currently provide 13 examples programs for Universal (resp. 5 for Python), and 11 configurations for Universal (resp. 6 for Python).

\section{Related Work}
\label{sec:rw}

In previous works, developers relied on server-side computations to provide web interfaces for static analyzers relying on relational numerical domains.
This includes for example Interproc \cite{DBLP:journals/sosym/Jeannet13,interproc-online}, Banal \cite{banal,banal-online} and FuncTion \cite{function,function-online}.
While client-server implementations require less work to adapt a static analyzer to the web, these implementations raise more security concerns and can be less scalable in the number of users.
In addition, these implementations need to maintain a compatible server service over the years.
This proves to be difficult: at the time of writing, the majority of these approaches are no longer operational.

More recently, pure client-side web interfaces have been developed for static analyzers.
These scale much better with the number of concurrent users, and their ``only'' dependency is a web browser -- one of the most widespread pieces of software.
\citet{salto,salto-online} provide a demo web interface for their Salto analyzer
of OCaml code, following previous demonstrators accompanying works around the
static analysis of functional languages
\cite{DBLP:journals/pacmpl/MontaguJ20,DBLP:conf/pldi/MontaguJ21}.
\citet{ciao,ciao-online} provide a WebAssembly version of Ciao Prolog, embedding the CiaoPP analyzer, which also supports interactive user inputs.
In their BINSEC tutorial at PLDI 2025, \citet{binsec-online} provide a web tutorial where users can choose between running BINSEC \cite{binsec} in CLI or running analysis scenarios directly in their browser.
To the best of our knowledge, Try-Mopsa is the only web interface supporting the analysis of multiple languages (our toy imperative language and Python), numerical relational domains, and an interactive exploration of the analysis.

Note that another lightweight approach for trying out some new software is the use of Docker containers. We actually provide Docker containers with each release of Mopsa \cite{mopsa_docker_containers}; these containers supports all features of Mopsa (e.g., including the C analysis).

\citet{DBLP:conf/tfm/NegriniAOCF24} describe how they use their static analysis platform LiSA within their static analysis courses.
Students are tasked with installing LiSA, and implementing several simple abstract domains.
Try-Mopsa aims at providing a zero-install static analyzer relying on complex abstract domains -- such as polyhedra -- to illustrate how such a tool works.
Evaluating the impact of Try-Mopsa on real teaching cohorts is left as future work.

Following studies about developer use of static analyzers \cite{DBLP:conf/kbse/ChristakisB16,DBLP:journals/tse/DoWA22}, there are quite a few works discussing interfaces for static analyzers.
MagpieBridge \cite{DBLP:conf/ecoop/LuoDB19} aims at simplifying the display of analysis results within IDEs supporting the Language-Server Protocol.
Several interfaces have also been developed to debug static analyzers \cite{DBLP:journals/tse/DoKH0B20}, provide an abstract debugger \cite{DBLP:conf/onward/HolterH0SV24}, or mix concrete and abstract debuggers  \cite{DBLP:conf/sle/MolleVR23}.
\citet{DBLP:conf/sefm/Correnson22} recently introduced a new graphical interface for Frama-C users.

Outside of the abstract interpretation community, \citet{DBLP:conf/sose/KosmatovWBR13} provide a unit testing as a service platform.
Both the Alt-Ergo SMT Solver \cite{conchon:hal-01960203} and the Why3 deductive verification platform \cite{DBLP:conf/esop/FilliatreP13} leverage JSOO to provide zero-install web versions \cite{tryaltergo,trywhy3}.
\citet{DBLP:journals/pacmpl/CanouCH17} describe the technical components required to run an OCaml MOOC (Massive Open Online Course) who attracted thousands of learners.
It relies in particular on a pure client-side development environment avoiding installation woes students could otherwise face.
\citet{DBLP:journals/corr/AriasPJ17} also provide a pure JavaScript version for the Rocq proof assistant, which aims at improving literate programming formatting and reducing installation burdens, in an educational context.

\section{Conclusion}
\label{sec:conclusion}

This article introduced Try-Mopsa, a pure client-side implementation of Mopsa running in the browser or as a progressive web app.
Try-Mopsa supports all the main core components of Mopsa, including support for relational numerical domains and its abstract debugger.
Try-Mopsa is designed to work on a wide range of devices, from smartphones to desktops, thanks to a responsive design and continuous testing of the support of popular browsers and various resolutions.
We envision Try-Mopsa to be a convenient tool to demonstrate static analysis capabilities, for teaching purposes or more generally to attract new users.

\paragraph{Data Availability Statement.} This paper is accompanied by an artefact \cite{monat_2025_17157479}, which has been peer-reviewed as ``available, functional and reusable''.

\paragraph{Acknowledgments.}
We thank the anonymous VMCAI reviewers and the participants of \href{https://www.dagstuhl.de/en/seminars/seminar-calendar/seminar-details/25421}{Dagstuhl Seminar \#25421} for their valuable feedback.
We are grateful to Aymane Ismail for prototyping an early version of Try-Mopsa; and to Alexandre Maréchal for his help in using the VPL. We thank Aymeric Fromherz, Louis Gesbert, Vincent Botbol, Louis Rustenholz, Antoine Miné, Abdelraouf Ouadjaout and the whole Mopsa team for their helpful discussion and comments on this work.

\todo[inline,color=blue!30]{List available abstract domains? And reductions, and combinators... Filter compared to already enabled ones?}

\printbibliography[title=Bibliography]

\doclicenseThis

\end{document}